\documentstyle[aps]{revtex}
\begin{document}
\voffset=-2mm
\title{METAL-INSULATOR TRANSITION IN THE TWO-DIMENSIONAL HUBBARD MODEL 
AT HALF-FILLING WITH LIFETIME EFFECTS WITHIN THE MOMENT 
APPROACH.}
\author{\it J.J. Rodr\'{\i}guez-N\'u\~nez}
\address{Departamento de F\'{\i}sica - CCNE,\\
Universidade Federal de 
Santa Maria, CAMOBI,\\ 
97105 - 900 Santa Maria - RS, 
Brazil. \\e-m: jjrn@mfis.ccne.br}
\author{\it S. Schafroth}
\address{Physik-Institut der Universit\"at Z\"urich,\\ 
Winterthurerstrasse 190, CH-8057 Z\"urich, 
Switzerland\\e-m: schafrot@physik.unizu.ch}
\date{\today}
\maketitle

%
%     Abstract
%
\begin{abstract}
We explore the effect of the imaginary part of the 
self-energy, $Im\Sigma(\vec{k},\omega)$, having 
a single pole, $\Omega(\vec{k},\omega)$, with 
{\it spectral} weight, $\alpha(\vec{k})$, and 
quasi-particle lifetime, $\Gamma(\vec{k})$, on the 
density of states. We solve the set of parameters, 
 $\Omega(\vec{k},\omega$), $\alpha(\vec{k})$, and 
$\Gamma(\vec{k})$ by means of the moment approach (exact sum rules) 
of Nolting\cite{1}. Our choice for $\Sigma(k,\omega)$, 
satisfies the Kramers - 
Kronig relationship automatically. Due to our choice of 
the self - energy, the system is not a Fermi liquid for 
any value of the interaction, a result which is also 
true in the moment approach of Nolting without lifetime 
effects. By increasing the value of   
the local interaction, $U/W$, at half-filling ($\rho = 
1/2$), we go from a paramagnetic metal to a paramagnetic insulator, 
(Mott metal - insulator transition ($MMIT$))\cite{2} 
for values of $U/W$ of the 
order of $U/W \geq 1$ ($W$ is the band 
width)  which is in agreement with 
numerical results for finite lattices and 
for infinity dimensions ($D = \infty$).  
These results settle down the main weakness 
of the spherical approximation of Nolting\cite{1}:  
a finite gap for any finite value of 
the interaction, i.e., an insulator for 
any finite value of $U/W$. Lifetime effects are 
absolutely indispensable. Our scheme works better than the 
one of improving the narrowing band factor, $B(\vec{k})$, 
beyond the spherical approximation of Nolting\cite{1}.\\
\\
Pacs numbers: 74.20.-Fg, 74.10.-z, 74.60.-w, 74.72.-h
\end{abstract}


%    PACS
%
%     I. Introduction
%
\section{Introduction}\label{sec:intro}

	After the discovery of 
the high-$T_c$ materials\cite{3}, 
the study of correlations has gained interested due 
to the fact that there is the belief\cite{4} 
that the normal properties of these materials could 
be explained in the framework of the Hubbard 
model\cite{5,6}, since electron correlations 
are strong, i.e., the on-site electron-electron 
repulsions $U$ are much larger than the energies 
associated with the hybridization of atomic orbitals 
belonging to different atoms\cite{7}. The strategy,  
according to Anderson\cite{8} to address the problem 
of the high-$T_c$ superconductivity, is to try to 
find a theory accounting for the normal state properties 
of the cuprates and then find an electron mechanism 
which destabilizes the normal state towards a 
superconducting state. The Hubbard Hamiltonian is a kind 
of minimum model\cite{9} which takes into account quantum 
mechanical motion of electrons in a solid, and nonlinear repulsion 
between electrons. Even thought this model is too simple to 
describe solids faithfully, serious theoretical studies have 
revealed that to understand the various properties of it is a 
very difficult task, since is the simplest many-body Hamiltonian
one can write down and which cannot be reduced to a single-particle
theory\cite{Assa}. Its study will prove useful in developing 
various notions and techniques in statistical physics of 
many degrees of freedom. Besides cuprate superconductors, the
Hubbard model may also be applicable to describing the metal -
insulating transition in materials like $La_xSr_{1-x}TiO_3$ and
$V_{2-y}O_3$ in which paramagnetic metal, antiferromagnetic
insulator, and a phase antiferromagnetic
metal are clearly separated in temperature -
pressure phase diagram\cite{all}. All these features, found in the
Hubbard model, make it suitable to reproduce experimental data.
Additionally, the study of correlations by itself in the Hubbard 
model is a rewarding task since it will shed light on still unsolved points of
the novel materials. For example, at high temperatures ($T_c~30-130~K$) these
HTSC cuprates, which are poor conductors, become superconductors. 
This feature is strange indeed because the Coulomb repulsion is
strong. Furthermore, the behavior of these materials at $T~>~T_c$ is even 
more puzzling than the superconductivity itself. Contrary to the 
predictions of the Fermi liquid theory, the resistivity at $T~>~T_c$ 
and optimum doping is linear in temperature, i.e., $R~\approx~T$. This suggests a very strong
scattering of elementary excitations. A discussion of the possible 
breakdown of Fermi liquid theory is given in Ref.\cite{Alexei}. There is also another approach which 
consists in studying toy models, i.e., exactly solvable,\cite{toy} 
in order to get some ideas on the one-particle properties of highly correlated electron systems.

	In this Letter, we will use the moment approach 
(or sum rules) of Nolting\cite{1} for the spectral density,  
$A(\vec{k},\omega)$. As it is well known in the literature\cite{ours}, one 
of the drawbacks of the moment approach in the spherical approximation - where 
$B(\vec{k})$, the narrowing band factor, is not $k$ - dependent - 
is that we {\it always}  find a gap in the density of states $(DOS)$. 
If the chemical potential happens to be in this gap, then we always have an insulator. 
It has been argued that the way to cure this unrealistic gap is 
to have a better approximation for the narrowing band factor, $B(\vec{k})$, 
task which has proven to be hard just to the present\cite{ours}. It is due to that 
we have followed a different path which consists in proposing a single pole 
structure in the self-energy, $\Sigma(\vec{k},\omega)$.  

    	The model we study is the  Hubbard Hamiltonian
\begin{eqnarray}\label{Ham}
H = t_{\vec{i},\vec{j}}
	c_{\vec{i}\sigma}^{\dagger}c_{\vec{j}\sigma}
   + \frac{U}{2} n_{\vec{i}\sigma}n_{\vec{i}\bar{\sigma}}   
   - \mu c^{\dagger}_{\vec{i}\sigma}c_{\vec{i}\sigma}~~,
\end{eqnarray}
where $c_{\vec{i}\sigma}^{\dagger}$ ($c_{\vec{i}\sigma}$) are creation
(annihilation) electron operators with spin $\sigma$. $n_{\vec{i}
\sigma} \equiv c_{\vec{i}\sigma}^{\dagger}c_{\vec{i}\sigma}$. 
$U$ is the local interaction, $\mu$ the chemical 
potential and we work in the grand canonical ensemble. We have 
adopted Einstein convention for repeated indices, i.e., for the $N_s$ 
sites $\vec{i}$, the $z$ nearest-neighbor sites $(n.n.)$  
$\vec{j}$ and for spin up and down ($\sigma = -\bar{\sigma} 
= \pm 1$). $t_{\vec{i},\vec{j}} = -t$, for $n.n.$ and zero otherwise.

	Let us propose for the self - energy, 
$\Sigma(\vec{k},\omega)$, the following single pole Ansatz
\begin{equation}\label{single}
\Sigma(\vec{k},\omega) = \frac{\alpha(\vec{k})}{\omega - 
\Omega(\vec{k}) - i\Gamma(\vec{k})}~~~ ; ~~~ 
\alpha(\vec{k}),\Gamma(\vec{k}) \in \Re ~~~   
\end{equation}
With our choice for $\Sigma(\vec{k},\omega)$, we have the real 
and imaginary parts of the self - energy satisfy the 
Kramers - Kronig relations\cite{CL}.

	By definition the one-particle Green function, 
$G(\vec{k},\omega)$, in terms of $\Sigma(\vec{k},\omega)$ is given as
\begin{equation}\label{1PGF}
G(\vec{k},\omega)  = \frac{1}{\omega - \varepsilon_{\vec{k}} 
- \Sigma(\vec{k},\omega)} ~~~ ,
\end{equation}
\noindent where $\varepsilon_{\vec{k}} = - 2t (\cos(k_x) 
+ \cos(k_y)  - \mu$. Also, we will require the spectral density, 
$A(\vec{k},\omega)$ which is defined as
\begin{equation}\label{Akw}
A(\vec{k},\omega) = - \frac{1}{\pi} 
\lim_{\delta \rightarrow 0^+} Im 
G(\vec{k},\omega + i\delta) ~~~ 
\end{equation}

	Using Eqs. (\ref{single} - \ref{Akw}), we arrive to the 
following expression for the spectral density
\begin{equation}\label{11}
A(\vec{k},\omega) = \frac{-1}{\pi} 
\frac{\alpha(\vec{k}) \Gamma(\vec{k})} 
{\left( (\omega - \varepsilon_{\vec{k}})(\omega - 
\Omega_{\vec{k}}) - \alpha(\vec{k}) \right)^2 + 
\Gamma^2(\vec{k})(\omega - \varepsilon_{\vec{k}})^2 }
\end{equation}

	Using the first three sum rules of  Nolting\cite{1} 
for the the spectral function of Eq. (\ref{Akw}) we obtain 
the following equations
\begin{eqnarray}\label{3mom's}
\int_{-\infty}^{+\infty} A(\vec{k},\omega)d\omega &=& 
a_o(\vec{k}) ~~~ \nonumber \\
\int_{-\infty}^{+\infty} \omega A(\vec{k},\omega)d\omega  
&=& a_1(\vec{k}) ~~~
\nonumber \\
\int_{-\infty}^{+\infty} \omega^2 A(\vec{k},\omega)d\omega  
&=& a_2(\vec{k})  
~~~ , 
\end{eqnarray}
\noindent where 
the $a_i(\vec{k})$'s, 
$i=0,2$ are given 
in Ref.\cite{1} (see also Ref.\cite{ours}). 
We do not use the forth moment 
or sum rule because we are at half - filling. Then, 
we assume that at $\rho = 1/2$ the chemical 
potential, $\mu = U/2$. This is verified as we will see 
from our results. At this moment, we would like 
to say once more that the drawback of previous calculations 
(included ours) consists in that the density of states which results of 
the two pole Ansatz for the one - particle 
Green function, in the spherical 
approximation of Nolting\cite{1}, always has a gap. This 
solution (always a gap) is known in the literature as 
the Hubbard-I solution\cite{IS} which has been critized since 
many years ago by Laura Roth\cite{LR} among others. We 
call the attention to Ref.\cite{oles} where the authors point out to the 
fact that the $\vec{k}$-dependence has to be included in 
$B(\vec{k})$. However, they study the 
negative Hubbard model in the strong coupling 
limit and in this limit we have a well developed correlation 
gap, anyway. Ole\'s 
and coworkers are practicioners of 
the two - pole ansatz for the one - particle Green function, 
without lifetime effects. Here we are including lifetime effects 
as a crucial ingredient in the formulation. Morover, our 
results clearly show that the cure the Hubbard-I solution in the 
moment approach is not in the $B(\vec{k})$ - term.

	In Fig. 1 we present the {\it energy spectrum} 
of the self - energy for several values of the Brillouin zone, 
for $U/t = 4$. In Fig. 2 we show the $\vec{k}$-dependence 
of $\alpha(\vec{k})$ for $U/t = 4, 6, 8, 12$. In Fig. 3 we present 
the $\vec{k}$ dependence of $\Gamma(\vec{k})$ for several values 
of the Brillouin zone, for  $U/t = 4, 6, 8, 12$. In Fig. 4 we 
show the density of states, $N(\omega)$ vs $\omega$. $N(\omega)$ 
is defined as
\begin{equation}\label{dos}
N(\omega)  = \frac{1}{N_s} \sum_{\vec{k}} A(\vec{k},\omega) ~~~
\end{equation}
\noindent In all the figures we have worked with a square lattice 
with periodicity of $N_s = 32 \times 32$. We conclude that the 
{\it spectral weight}, $\alpha(\vec{k})$ and the damping factor, 
$\Gamma(\vec{k})$, do not strongly depend on $\vec{k}$, 
for small values of $U/W$. For example, 
for $U/t = 4.0$, $\Gamma(\vec{k})/t \approx 1.5$. 
However, for larger values of $U/W$, $\alpha(\vec{k})$ has some 
$\vec{k}$ dependence. On the contrary, $\Gamma(\vec{k})$ shows a 
stronger $\vec{k}$ - dependence for larger values of $U/W$. We 
also note that $\alpha(\vec{k}) \times \Gamma(\vec{k}) < 0$ 
in order to give a positive spectral density (see Eq.(\ref{11})). 
Taking a look to 
the density of states, $N(\omega)$ vs $\omega$, we see 
that the correlation gap opens up 
for $U/W \geq 1.0$. This is equivalent to the Hubbard-III like 
solution\cite{III}. Due to our choice of the self - energy, we do not 
have a Fermi liquid\cite{8,III}. Edwards and Hertz\cite{EH} have 
studied the breakdown of Fermi liquid theory in the Hubbard model 
at $T = 0$. They have drawn a phase boundary between a Fermi liquid 
and a non - Fermi liquid ($\rho~vs~U$ in the 
paramagnetic phase). According to these authors 
for small values of $U$ we have a Fermi liquid behavior 
(metallic Fermi liquid) and for larger values of $U$ we have 
a non - Fermi liquid metal. In our approach we have a non - 
Fermi liquid behavior for any value of interaction. In order to 
agree with the authors of reference\cite{EH} we have to include 
a $\omega^2$ behavior for frequencies close by to the chemical 
potential in addition to the single pole structure chosen in the 
present work. A metallic Fermi to metallic non - Fermi liquid 
behavior has been obtained by Figueira, Anda and Nogueira\cite{b}. 
We should mention that the phase diagram of the Hubbard model 
contains the anti - ferromagnetic transition\cite{40} which has not 
been considered here.

	In short, we have postulated a one - pole Ansatz for the 
self - energy (Eq. (\ref{single})) in the moment approach of 
Nolting\cite{1}. Our essential new working idea 
with respect to the now canonical method of Nolting is that we 
start from the self - energy while Nolting proposes a two pole 
ansatz for the one - particle Green function from the begining. 
In both approaches sum rules for the spectral function, 
$A(\vec{k},\omega)$, are imposed. 
As function of interaction, we see a 
Mott insulator - transition (MMIT) from a metal at weak 
interaction (no gap at the zero frequency) to a well developed 
gap (an insulator) for large values of the local interaction. 
Our formulation is not equivalent to the infinite dimension 
calculation of Georges et al\cite{a} since we have included the 
$\vec{k}$ - dependence of the selfenergy (we are working in two 
dimensions). There is the calculation of Figueira, Anda and 
Nogueira\cite{b} where the authors have also neglected the $\vec{k}$ - 
dependence of $\Sigma(\vec{k},\omega)$. In particular, the 
authors of Refs.\cite{b,a} find a Kondo peak at $\omega = 0$. In 
our case, $\rho = 1/2$, the peak-like structure seen at the 
chemical potential is most likely due to the van Hove logarithmic 
singularity which always goes away with the opening of the correlation 
gap. With our precision\cite{201}, we cannot conclude about the 
presence or not of the Kondo peak. Nolting 
himself\cite{c} has also studied the effect of damping on 
magnetism. We leave for the future the comparison with Ref.\cite{d} where 
the authors discuss the use of the moment approach for interpolating 
$\Sigma(\vec{k},\omega)$ between weak and strong interaction. We include 
Figure 5 to present $\Omega_{(m,n)}$ vs $n$ for fixed values of 
$m$, for $U/t = 4.0$ and a mesh of $10 \times 10$. It is worth 
looking to the band structure for $m = 0$. This curve has 
quasi the same structure than the band structure shown in Figure 
1, for a bigger mesh ($32 \times 32$). Then, we can say that finite 
size effects are minimal at least with respect to the band 
structure calculations. (See the discussion of Ref.\cite{201}). 
Finally, we would like to say that the non Fermi liquid behavior 
that we have in our approach (by construction) has nothing to do 
with any microscopic mechanism. Our arguments rely on general 
grounds of a self - energy ansatz which is used in the one - particle 
Green function to which exact sum rules (moments) are imposed. 
Those interested in physical mechanisms of non - Fermi liquid 
behavior are encouraged to take a look to the nice account presented 
in Anderson's book\cite{TT}.

    We would like to thank  CONICIT (project F-139) and 
the Brazilian Agency CNPq (project 300705/95-96) for finantial support.  
Interesting discussions with Prof. H. Beck, Prof. M.S. Figueira, Prof. E. 
Anda and Dr. 
M.H. Pedersen are fully acknowledged. We thank Mar\'{\i}a Dolores 
Garc\'{\i}a Gonz\'alez for a reading of the manuscript. 
%
%     Acknowledgments
%
%
%     References.
%     ^^^^^^^^^^^
%

\vspace{1.6cm}
\begin{center}
{\huge FIGURES}
\end{center}

\vspace{0.8cm}
\noindent
Figure 1. $\Omega_{0,n}$ vs n for 
$U/t = 4.0$ and $\rho = 0.5$. The $\vec{k}$ vector 
is given by $\vec{k} = \pi(0,n)/32$. As we work in 
two dimensions, the bandwidth is $W = 8t$. Compare with 
$m = 0$ of Figure 5.

\vspace{0.6cm}
\noindent 
Figure 2. $\alpha(0,n)~vs~n$ for different values of $U/t$. 
Same parameters as in Figure 1.

\vspace{0.6cm}
\noindent  
Figure 3. We show $\Gamma(0,n)~vs~n$ for different values of $U/t$. 
Same parameters as in Figure 1.

\vspace{0.6cm}
\noindent Figure 4. $N(\omega)~vs~\omega$ for 
$U/t = 4, ~12$. We see the opening of the correlation gap for 
$U/W \geq 1.0$, signaling the Mott metal - insulator transition.

\vspace{0.6cm}
\noindent Figure 5. $\Omega_{m,n}$ vs $n$ for some fixed 
values of $m$. $U/t = 4.0$, $\rho = 0.5$ and our mesh is 
$10 \times 10$. Compare $m = 0$ with Figure 1.

\begin{references}
\bibitem{201}
	Our calculations have been done for a finite 
	$\vec{k}$ - mesh of $32 \times 32$. Most likely we 
	have to consider larger meshes. However, the main 
	conclusion of the present paper will remain: 
	the Mott metal - insulator transition ($MMIT$).
\end{references}
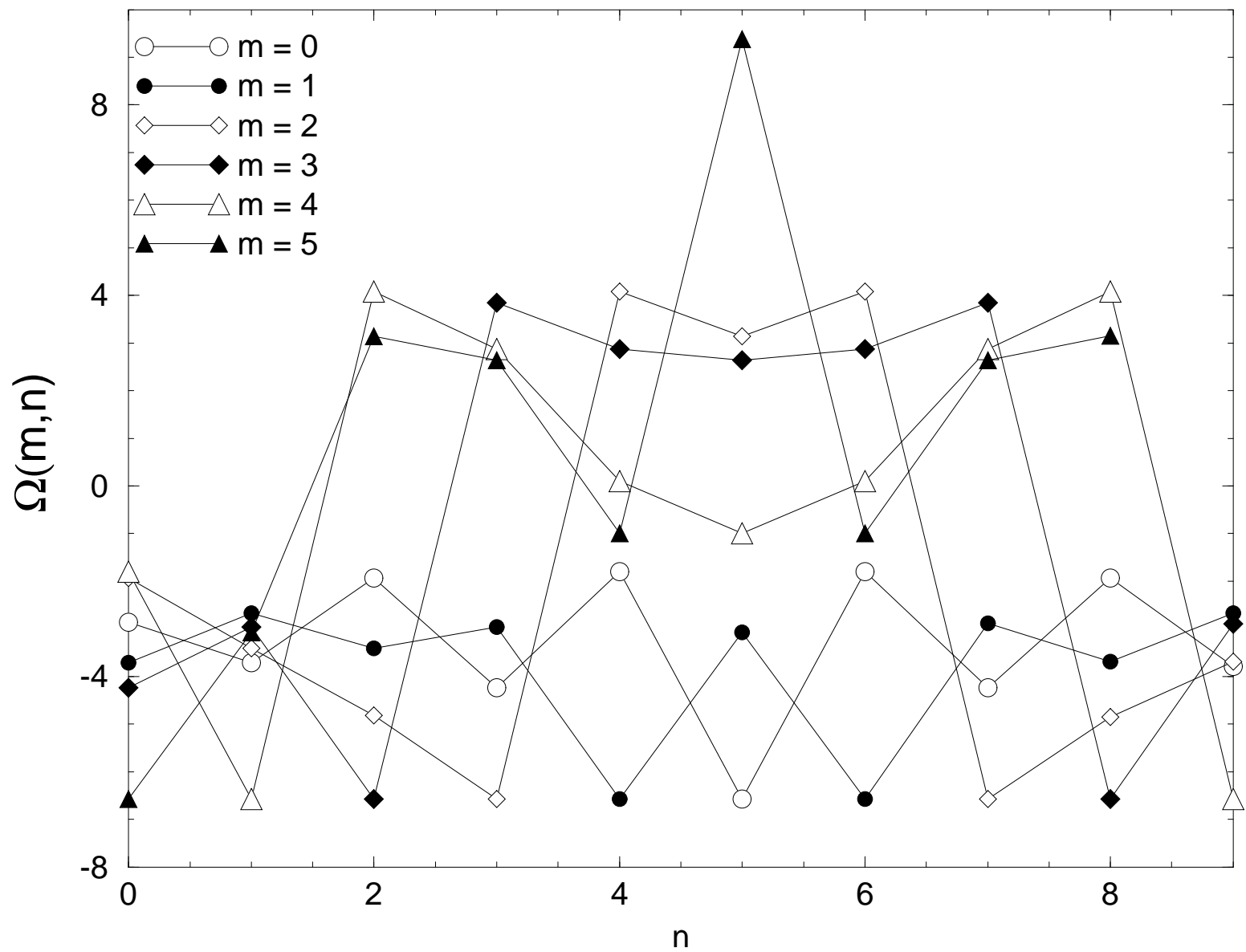
\end{document}